# Tracking multiscale liquid-crystal order with high-harmonic spectroscopy

Marta L. Murillo-Sánchez[1†], Andrea Annunziata[1†],
Simon Vendelbo Bylling Jensen[2], Ángel Rubio[2,3],
Caterina Vozzi[4], Thomas Pfeifer[1], Davide Faccialà[4],
Nicolas Tancogne-Dejean[2,5*], Laura Cattaneo[1*]

[1]Max-Planck-Institut für Kernphysik, Heidelberg, Germany

[2]Max Planck Institute for the Structure and Dynamics of Matter and Center for Free-Electron Laser Science, Hamburg, Germany

[3]Initiative for Computational Catalysis (ICC), The Flatiron Institute, New York, New York, USA

[4]CNR, Istituto di Fotonica e Nanotecnologie (CNR-IFN), Milano, Italy

[5]Research Center Future Energy Materials and Systems of the University Alliance Ruhr and Interdisciplinary Centre for Advanced Materials Simulation, Faculty of Physics and Astronomy, Ruhr University Bochum, Bochum, Germany

*Corresponding author. Email: nicolas.tancogne-dejean@ruhr-uni-bochum.de; laura.cattaneo@mpi-hd.mpg.de

†These authors contributed equally to this work.

Can high-harmonic generation track how microscopic intermolecular interactions give rise to macroscopic order? We address this question in the thermotropic liquid crystal 4-cyano-4'-*n*-pentylbiphenyl, a model system with a well-defined nematic-to-isotropic transition. Using a temperature-controlled flat jet, we show that polarization-resolved high-harmonic generation driven by mid-infrared strong-field pulses is highly sensitive to liquid-crystal order. In the nematic phase, the harmonic yield forms a four-lobed angular pattern with twofold rotational symmetry and is strongest when the driving-field polarization is perpendicular to the molecular director; this anisotropy vanishes in the isotropic liquid. A tight-binding model parameterized by first-principles calculations attributes this response to intermolecular orbital hybridization in short-range antiparallel molecular arrangements. Temperature-dependent measurements across multiple harmonic orders resolve the same first-order phase transition, with each harmonic order exhibiting a distinct optical sensitivity. A multiscale model combining the microscopic response, Landau–de Gennes orientational correlations, and phase matching explains this behaviour through the interplay between orientational correlation lengths and harmonic coherence lengths. High-harmonic generation therefore links local electronic coupling to mesoscopic order and macroscopic propagation, providing an all-optical probe of multiscale organization in soft matter.

# 1 Introduction

High-harmonic generation (HHG) is a powerful tool for producing coherent extreme-ultraviolet (XUV) radiation (*1–4*). Beyond serving as a light source, HHG provides access to electronic structure and ultrafast dynamics on subcycle timescales and with attosecond temporal resolution, making it a sensitive spectroscopic probe of matter.

In gases, HHG is commonly described by the three-step model (*5, 6*). This model comprises tunnel ionization, laser-driven acceleration, and recombination of the returning electron wave packet with its parent ion, thereby encoding information about the electronic structure of the target in the emitted harmonics. This mechanism underpins applications of high-harmonic spectroscopy (HHGS), including molecular imaging and orbital tomography (*7*), as well as the tracking of coupled electron–nuclear dynamics in systems ranging from diatomic to complex polyatomic molecules (*8–12*).

In solids, HHGS has enabled the all-optical reconstruction of band dispersions under both equilibrium (*13, 14*) and nonequilibrium conditions (*15*), the measurement of petahertz intraband currents (*16*), and access to symmetry-dependent electronic properties (*17, 18*), Berry phases (*19*), and other band-structure singularities (*20*). Polarization-resolved and chiral HHGS measurements have further demonstrated that harmonic emission encodes information about crystal symmetry and collective lattice dynamics (*21, 22*), including intermolecular electronic coupling in organic molecular systems (*23*).

In liquids, the absence of long-range order complicates the extraction of structural information from a structurally disordered ensemble. Nevertheless, HHG enables field-driven electrons to probe their local environment through a recollision process adapted to the condensed phase (*24*). Specific spectral features can act as fingerprints of the microscopic environment (*25, 26*), with reported sensitivity extending even to the secondary solvation shell of a liquid (*27, 28*). HHGS can therefore retain pronounced sensitivity to local structure and electronic configuration even in the absence of long-range order.

A key question remains open: Can HHG quantitatively track a thermodynamic order parameter as it evolves through a structural phase transition? In particular, can it resolve the gradual temperature dependence of partial orientational order and its discontinuous loss at a weakly first-order transition? Addressing this question is crucial because it would provide access not only to structural evolution but also to the associated reorganization of electronic structure and its coupling to nuclear degrees of freedom, as demonstrated in simpler molecular systems (*8–12*). Establishing HHG as a probe of such transitions would open an all-optical window onto how microscopic electronic interactions evolve as macroscopic order changes, thereby bridging local interactions and thermodynamic phases.

Thermotropic liquid crystals (LCs), with their rich sequence of mesophases between the crystalline solid and isotropic liquid, provide an ideal testbed for this question. Beyond their established applications, LCs are archetypal dynamically disordered systems and are relevant to flexible electronics, organic photovoltaics, molecular sensors, and neuromorphic computing, technologies that rely on molecular configurations and interactions that fluctuate in space and time (*29–33*). In nematic-forming LCs, heating beyond the crystalline melting point removes positional order while preserving long-range orientational alignment. This defines the nematic ($N$) phase, in which the molecules align, on average, along a director $\hat{\mathbf{n}}$ (Figure 1a). Further heating eliminates macroscopic orientational order through a weakly first-order phase transition to an isotropic ($I$) liquid (Figure 1c), although short-range nematic correlations may persist close to the transition (Figure 1b).

Here, we use HHGS to track the nematic-to-isotropic ($N$–$I$) transition. Several temperature-dependent quantities describe this transition, among them the scalar orientation order parameter $S$, which quantifies the average molecular alignment relative to $\hat{\mathbf{n}}$. Here, $S = 1$ corresponds to perfect alignment, whereas a macroscopically isotropic orientational distribution corresponds to $S = 0$. Importantly, the vanishing of the macroscopic order parameter does not preclude finite-range local orientational correlations in the isotropic phase. Describing the phase transition therefore also requires accounting for pretransitional fluctuations and finite-range correlations through two characteristic correlation lengths (Figure 1d). Within the effective description adopted here, the first, $\xi_{\mathrm{nem}}$, characterizes the spatial scale over which orientational fluctuations remain correlated in the nematic phase near the transition. These fluctuations include variations in both the magnitude of the order parameter, $\delta S$, and the director orientation, $\delta\mathbf{n}$ (*34*) (Figure 1d, solid orange curve).

The second correlation length, $\xi_{\mathrm{iso}}$, characterizes the spatial extent of residual nematic correlations in the isotropic phase above the transition, often described as *paranematic swarms* (Figure 1b). Their characteristic size decreases with increasing temperature (*35*), ultimately approaching the intermolecular length scale, $\xi_0 \approx 4\,\mathrm{\AA}$, a few tens of kelvins above $T_{\mathrm{NI}}$ (Figure 1d, red-orange dash-dotted curve). Together, these parameters highlight the multiscale and statistical complexity of the $N$–$I$ phase transition.

The hierarchy of spatial scales underlying the $N$–$I$ phase transition in cyanobiphenyl LCs has traditionally been investigated using transient-grating and optical-Kerr-effect spectroscopies (*36, 37*). These all-optical techniques are sensitive to pretransitional orientational dynamics (*38–42*), but they primarily probe low-order, ensemble-averaged optical responses and are only indirectly sensitive to changes in the underlying electronic structure. This limitation motivates the use of a complementary, strongly nonlinear probe such as HHG, whose dependence on the molecular orientation and local electronic structure can provide access to the microscopic electronic rearrangements accompanying the phase transition.

Building on our recent demonstration of polarization-resolved high-harmonic spectroscopy in an LC confined within static cells (*43*), we now implement this approach in a temperature-controlled, recirculating flat jet (*44*). The jet continuously renews the sample, reducing cumulative laser-induced degradation and thermal loading, while hydrodynamic shear aligns the molecules without an external electric or magnetic field. This configuration enables systematic, high-repetition-rate measurements of the temperature- and alignment-dependent harmonic response of the prototypical thermotropic LC 4-cyano-4′-*n*-pentylbiphenyl (5CB) (inset of Figure 1d).

We show that HHGS quantitatively tracks optical signatures of the structural and electronic evolution across the $N$–$I$ phase transition. The high-harmonic yield exhibits a pronounced, orientation-dependent nonlinear anisotropy arising from the interplay between intra- and intermolecular electronic couplings. This sensitivity reveals that field-driven electronic dynamics probe the local molecular environment and capture orientation-dependent signatures of intermolecular hybridization in the nematic phase. In the isotropic fluid, local intermolecular interactions may persist, but their anisotropic contributions become rotationally averaged in the macroscopic harmonic response.

To interpret these signatures, we develop a first-principles-informed multiscale framework that extends the approach introduced in Ref. (*45*). This framework links subcycle microscopic electron dynamics to macroscopic propagation, phase matching and thermodynamic quantities such as the orientational order parameter $S$ and the characteristic correlation lengths of the two phases. This framework thereby extends the information accessible through conventional low-order nonlinear optics and provides an all-optical route for resolving mesoscopic structural evolution and order–disorder transitions in soft matter.

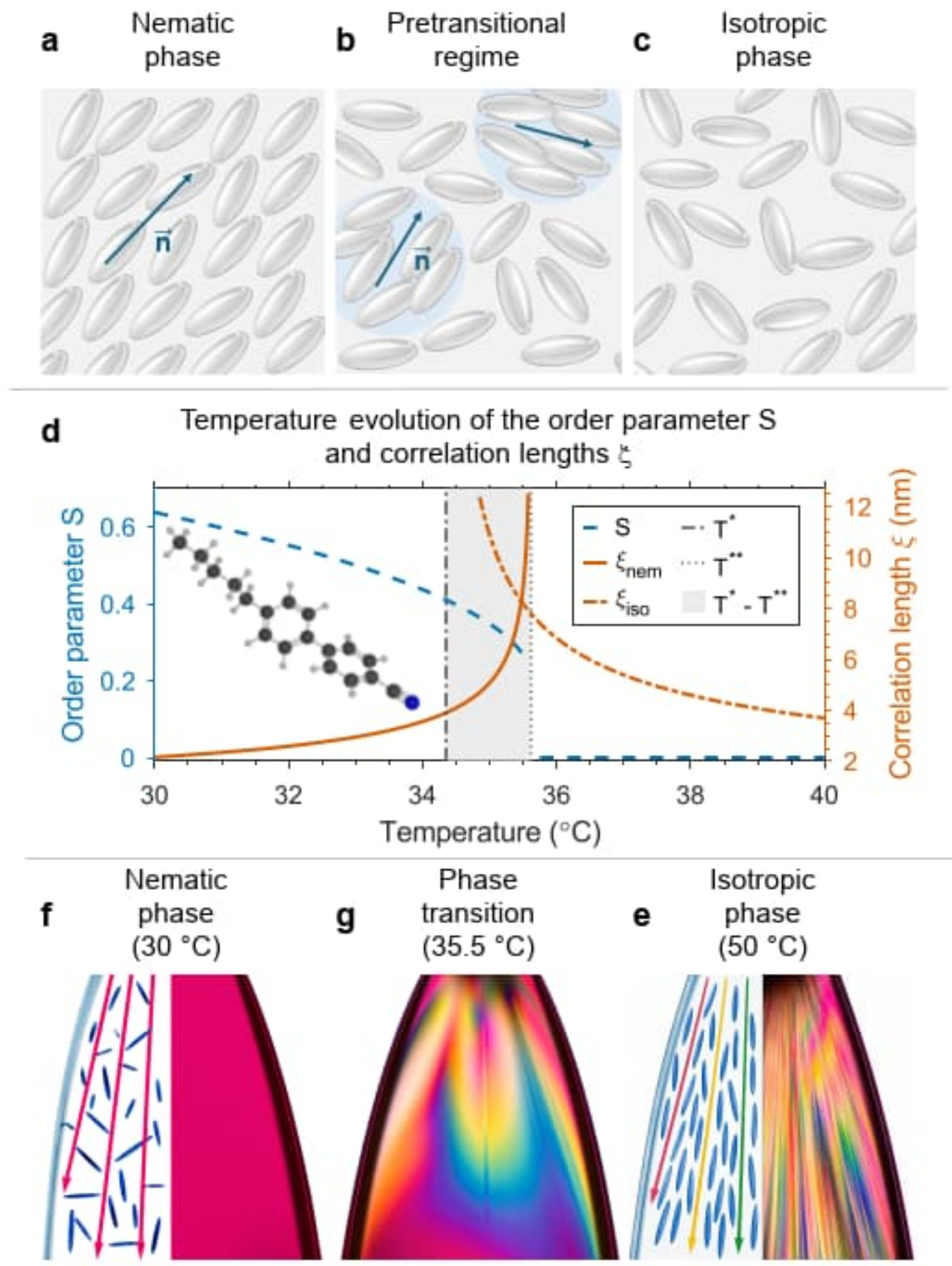

**Figure 1**: **Nematic–isotropic phase transition in an LC flat jet. a–c,** (**a**) Schematic molecular organization in the nematic phase; (**b**) the pretransitional regime with persistent short-range orientational correlations; (**c**) the isotropic phase. **d,** Temperature evolution of the orientational order parameter $S$ (blue dashed curve, left axis) and the nematic and isotropic correlation lengths, $\xi_{\mathrm{nem}}$ and $\xi_{\mathrm{iso}}$ (orange solid and dash-dotted curves, respectively; right axis). The correlation lengths describe the spatial extent of orientationally correlated regions below and above the transition. Vertical dash-dotted and dotted lines indicate the supercooling and superheating limits, $T^{*}$ and $T^{**}$, respectively, which delimit the metastability interval of the first-order transition. The shaded region marks the interval $T^{*} < T < T^{**}$. The inset shows the molecular structure of 5CB. **e–g,** Polarized optical micrographs of the flowing 5CB flat jet in the nematic phase at 30 °C (**e**), near the phase transition at 35.5 °C (**f**) and in the isotropic phase at 50 °C (**g**). Schematics in **e** and **g** illustrate the corresponding flow-induced molecular organization.

## Results

We investigated 5CB, a prototypical thermotropic LC whose vapour pressure, viscosity and refractive index are well characterized over broad temperature and spectral ranges (*46–48*). The sample was delivered as a temperature-controlled, recirculating flat jet formed by the symmetric collision of two identical jets, producing a planar sheet with a thickness of $23.1 \pm 0.2\,\mu$m. Temperature stabilization enabled measurements across the $N$–$I$ transition within the same flowing sample. Molecular alignment was generated solely by hydrodynamic shear, without an externally applied electric or magnetic field (*35, 49–52*).

High harmonics were generated by intense mid-infrared pulses centred at $3.9\,\mu$m, with a duration of 130 fs and a peak intensity of approximately $18\,\mathrm{TW\,cm^{-2}}$. The pulses were produced by difference-frequency generation in an optical parametric amplifier pumped by a femtosecond Ti:sapphire laser and were focused onto the LC jet (Figure 2a). The emitted harmonics were detected using commercial spectrometers and measured as functions of driving-field polarization, intensity and sample temperature.

We independently characterized the mesophase and flow-induced molecular alignment using polarized optical microscopy with a full-wave ($\lambda$) retardation plate, which converts variations in birefringence into distinct interference colours (Supplementary Information). Figure 1e–g shows representative micrographs and corresponding molecular-alignment schematics for the nematic phase at 30 °C ($S = 0.5$), the near-transition regime at 35.5 °C, and the isotropic phase at 50 °C ($S = 0$), respectively.

## Intermolecular origin of the emission anisotropy

We obtained polarization-resolved angular distributions by recording HHG spectra as a function of the driving-field polarization in the nematic and isotropic phases and integrating each harmonic over its spectral bandwidth. In the nematic phase, the yield exhibits a pronounced four-lobed angular distribution (Figure 2b, filled circles). The strongest lobes occur when the driving-field polarization is perpendicular to the director $\hat{\mathbf{n}}$, whereas weaker lobes appear when the polarization is parallel to $\hat{\mathbf{n}}$, as indicated in Figure 2b,c. By contrast, the isotropic phase shows an angle-independent response (Figure 2b, open circles), consistent with rotational averaging over the disordered ensemble. The nematic response is consistent with the $D_{\infty h}$ symmetry of a uniaxial nematic (*35*) and with previous HHG measurements in static LC cells (*43*).

Spatially resolved measurements show that the four-lobed anisotropy is preserved across the flat jet, while its principal axes follow the local 5CB director (Figure 2c). Polarization-resolved HHGS can therefore map spatial variations in director orientation.

An isolated-molecule picture predicts a response dominated by fields polarized along the molecular long axis, corresponding to polarization parallel to the director. The dominant perpendicular emission therefore points to an additional intermolecular contribution. In 5CB, nematic order favours antiparallel molecular arrangements and close lateral packing, configurations associated with UV-induced excimer formation (*53–55*). Related $\pi$-stacked structures have also been proposed as ground-state precursors (*56*). Such preorganized configurations provide a plausible microscopic origin for the enhanced perpendicular emission, analogous to organic crystals in which intermolecular coupling governs the angular dependence of HHG (*23*).

Power-dependent measurements provide further evidence for distinct parallel and perpendicular emission channels (Figure 2d). The perpendicular response departs from its low-intensity power-law scaling at a lower intensity than the parallel response, most clearly for H7 and H9. The fitted intensity intervals and scaling exponents are reported in Supplementary Section 2.4.

To investigate the microscopic origin of this anisotropy, we performed density functional theory (DFT) calculations for 5CB monomers and dimers. Among the geometries considered, the antiparallel dimer is energetically favoured in the gas phase and exhibits intermolecular orbital hybridization and a geometry-dependent strong-field response consistent with the experimental observations (Supplementary Section 5). Upon dimerization, the lowest unoccupied molecular orbital becomes delocalized over the molecular pair, with substantial density extending along the intermolecular direction. This redistribution provides a microscopic basis for the enhanced perpendic-ular response and resembles the intermolecular hybridization identified in pentacene crystals (*23*).

We extended this picture to the condensed phase using a semi-empirical tight-binding (TB) model parameterized by the electronic-structure calculations. Each molecule was represented by a two-site model, with collective effects introduced through nearest-neighbour hopping in structural ensembles constrained by the temperature-dependent order parameter $S(T)$ and density $\rho(T)$. In the nematic phase, close lateral packing enhances hybridization perpendicular to the director. Including this coupling reproduces the observed perpendicular-to-parallel yield hierarchy, whereas independent elongated emitters yield a response dominated by fields polarized parallel to the molecular axes (*57*). Together, the calculations show that anisotropic intermolecular hybridization reshapes the strong-field response of the nematic liquid.

## Multiscale response across the phase transition

Having established HHG as a phase-sensitive probe of 5CB, we used it to follow structural evolution across the *N–I* transition (Figure 3). We recorded complete polarization-resolved angular distributions from 31.0–39.0 °C, with a temperature stability of ±0.1 °C. The mesophase was verified by *in situ* polarization imaging before each acquisition (Figure 3, top row, and Supplementary Section **??**).

Below the transition, all harmonic orders exhibit the four-lobed angular pattern. Between 34.7 and 35.3 °C, the H3 response becomes nearly isotropic, whereas H5, H7 and H9 retain pronounced angular contrast. By 35.8 °C, all orders are nearly isotropic. Higher-order harmonics therefore retain sensitivity to orientational anisotropy closer to the phase boundary than H3. The principal axis of the anisotropy also rotates with temperature (Supplementary Section **??**), consistent with a temperature-dependent balance between flow-induced alignment and nematic viscous response.

To quantify the polarization anisotropy of the HHG response, we define the polarization ratio

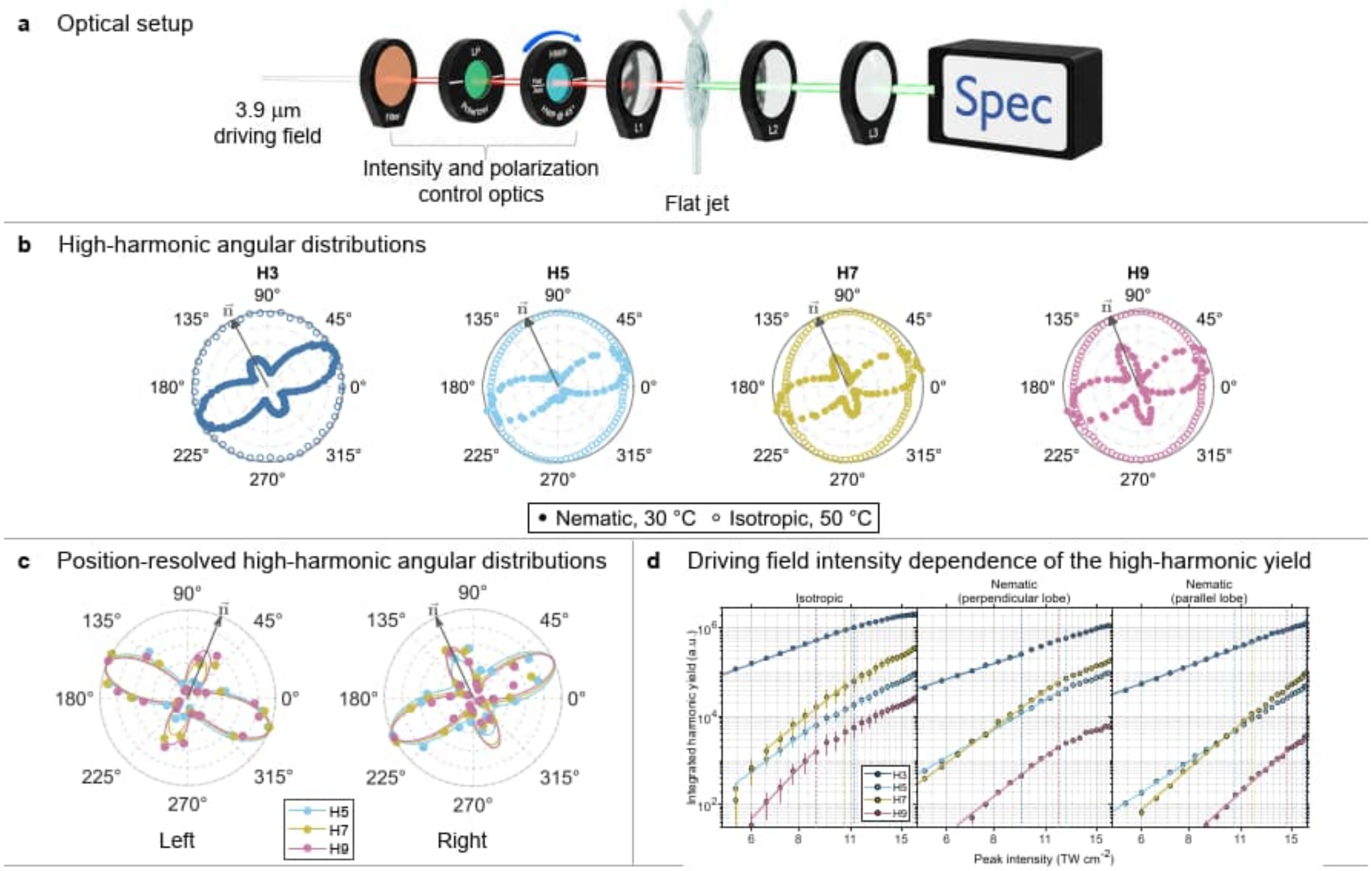

**Figure 2**: **Polarization-resolved high-harmonic response of flowing 5CB. a,** Experimental setup. Mid-infrared pulses centred at 3.9 $\mu$m are focused onto a temperature-controlled, recirculating 5CB flat jet. Wire-grid polarizers and wave plates control the peak intensity and polarization of the driving field, and the emitted harmonics are recorded spectrally. **b,** Angular distributions of H3, H5, H7 and H9 in the nematic phase at 30 °C (filled symbols) and the isotropic phase at 50 °C (open symbols). The arrows indicate the local nematic director $\hat{\mathbf{n}}$, aligned with the parallel-polarization axis. **c,** Position-resolved angular distributions of H5, H7 and H9 measured on the left and right sides of the flat jet, as viewed along the beam-propagation direction. **d,** Integrated harmonic yield as a function of the peak driving-field intensity in the isotropic phase and in the nematic phase for driving-field polarizations perpendicular and parallel to the director, shown from left to right. Open and filled symbols denote the isotropic and nematic phases, respectively. Solid lines show power-law fits, $I_{\mathrm{HHG}} \propto I_{\mathrm{drive}}^{n}$, over the intervals specified in Supplementary Section **??**. The extracted exponents are reported in Supplementary Table **??**.

$$R_q(T) = \frac{I_{\parallel}^{(q)}(T)}{I_{\perp}^{(q)}(T)}, \tag{1}$$

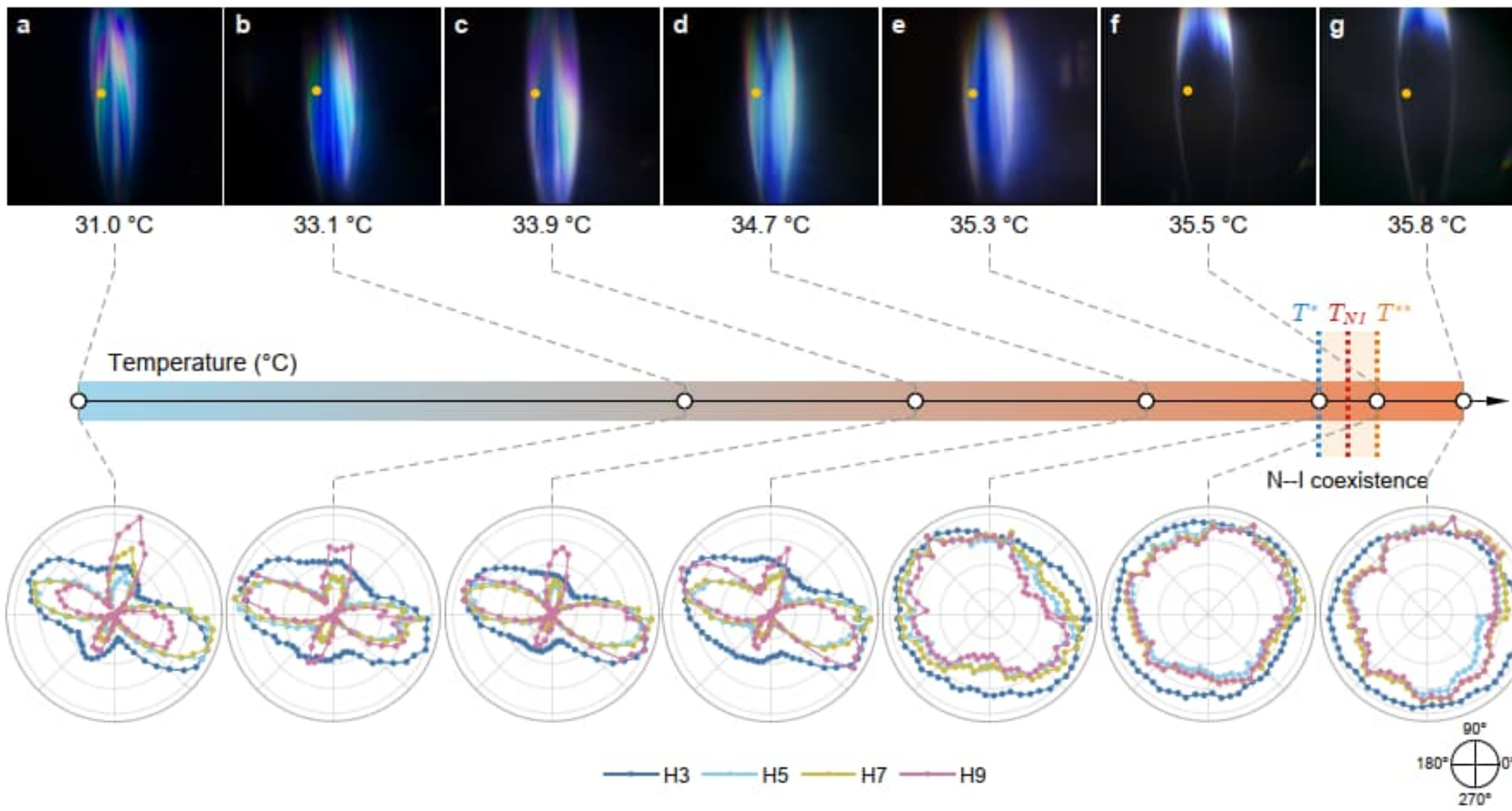

**Figure 3**: **Temperature-dependent high-harmonic response across the *N–I* transition. a–g,** *In situ* polarization images of the 5CB flat jet and corresponding angular distributions of H3, H5, H7 and H9 between 31.0 and 35.8 °C. The birefringence progressively vanishes on approaching the isotropic phase, whereas higher-order harmonics retain anisotropic emission to higher temperatures than H3. The arrow indicates increasing temperature. Dashed lines mark $T^*$, $T_{\mathrm{NI}}$ and $T^{**}$; $T^*$ and $T^{**}$ delimit the metastability interval of the *N–I* transition. The angular distribution measured in the isotropic phase at 39.0 °C was used as a calibration reference (Supplementary Information).

where $I_{\parallel}^{(q)}(T)$ and $I_{\perp}^{(q)}(T)$ are the intensities of harmonic order $q$ measured with the driving field polarized parallel and perpendicular to the local nematic director, respectively. Thus, $R_q(T) < 1$ denotes a dominant perpendicular response, whereas $R_q(T) = 1$ corresponds to an isotropic angular distribution.

Figure 4a shows the temperature dependence of $R_q(T)$ predicted by the TB model, with sigmoidal fits included as guides to the eye. The corresponding values extracted from the experimental angular scans are shown in Figure 4b (solid symbols). Both experiment and simulation show an abrupt change across the *N–I* transition, from a strongly anisotropic four-lobed response in the nematic phase to an almost isotropic yield in the isotropic phase. Near the transition, however, the microscopic TB model does not reproduce the measured temperature dependence of $R_q(T)$. For all harmonic orders, the experimental ratio decreases sharply as the transition is approached from the nematic side and rises above unity immediately above $T_{\mathrm{NI}}$. Both the amplitude of this feature and its characteristic temperature depend on harmonic order. This discrepancy highlights a central limitation of the microscopic model: its assumption of spatially uniform order parameter $S(T)$ and density $\rho(T)$. Although the model captures the competition between intra- and intermolecular emission away from the transition, it does not account for spatial phase coexistence, orientational fluctuations, or the finite correlation lengths that govern the response near $T_{\mathrm{NI}}$ (*35*).

To reduce systematic asymmetries, the two parallel maxima adjacent to each perpendicular maximum were each normalized by that perpendicular maximum, and the resulting ratios were averaged over repeated measurements. Error bars denote the standard error of the mean.

To capture the collective response near the phase boundary and model the experimental anisotropy ratios shown in Figure 4b, we complement the microscopic TB calculations with a macroscopic Landau–de Gennes $Q$-tensor description of the orientational order (*35*). The harmonic intensity is expressed as $I_q \propto \langle |E_q|^2 \rangle$, where $E_q$ is the field associated with the harmonic order $q$, obtained by averaging the TB-derived molecular response over the orientational distribution and coherently summing the field over the interaction volume (see Supplementary Information, Section 5.4). The model incorporates two classes of relevant mesoscopic length scales: the temperature-dependent orientational correlation lengths, $\xi_{\mathrm{nem}}(T)$ and $\xi_{\mathrm{iso}}(T)$, and the polarization-dependent harmonic coherence length, $L_{\mathrm{coh},q,\alpha}$, defined as

$$L_{\mathrm{coh},q,\alpha} = \frac{\pi}{|\Delta k_{q,\alpha}|}, \qquad \Delta k_{q,\alpha} = k_\alpha(q\omega) - q k_\alpha(\omega), \tag{2}$$

where $\alpha = \|, \perp$ denotes the driving-field polarization relative to the nematic director, and $\Delta k_{q,\alpha}$ is the corresponding phase mismatch between the generated $q$th-harmonic field and the $q$th-order nonlinear polarization driven by the fundamental field. The coherence length determines the distance over which the $q$th-harmonic field accumulates coherently (*58*). For H3–H9, $L_{\mathrm{coh},q,\alpha}$ ranges from tens to hundreds of nanometres, far exceeding both the intermolecular separation ($\sim 4$ Å) and the molecular length (~ 1 nm). The measured response is therefore governed not only by the local electronic structure but also by orientational correlations and phase-dependent propagation through the finite liquid sheet.

Figure 4c shows how the normalized phase mismatch evolves with temperature for H3 and H9. In the nematic phase, birefringence produces different phase mismatches for driving fields polarized parallel and perpendicular to the director. This polarization-dependent splitting is substantially larger for H9 than for H3 and decreases as the system approaches the isotropic phase, where the parallel and perpendicular responses converge to the isotropic reference. These phase-mismatch curves provide the propagation contribution used in the macroscopic fits in Figure 4b. Thus, the measured evolution of $R_q(T)$ reflects not only the loss of microscopic emission anisotropy shown in Figure 4a, but also the harmonic-dependent propagation efficiency of the two polarization channels.

We describe the evolution of $R_q(T)$ across the transition by combining the nematic and isotropic responses through a harmonic-dependent sigmoidal weight $W_q(T)$, which represents the progressive conversion between the two mesophases within the transition region:

$$R_q(T) = \frac{W_q(T) I_{\parallel}^{(q,\mathrm{nem})}(T) + \left[1 - W_q(T)\right] I^{(q,\mathrm{iso})}(T)}{W_q(T) I_{\perp}^{(q,\mathrm{nem})}(T) + \left[1 - W_q(T)\right] I^{(q,\mathrm{iso})}(T)}, \tag{3}$$

$$W_q(T) = \frac{1}{1 + \exp\left[(T - T_q)/\Delta T_q\right]}. \tag{4}$$

Here, $I_{\parallel}^{(q,\mathrm{nem})}(T)$ and $I_{\perp}^{(q,\mathrm{nem})}(T)$ denote the $q$th-harmonic intensities obtained for driving-field polarizations parallel and perpendicular to the nematic director, respectively, and $I^{(q,\mathrm{iso})}(T)$ is the isotropic-phase response. The parameter $T_q$ defines the optical-response midpoint for harmonic order $q$, and $\Delta T_q$ controls the crossover width.

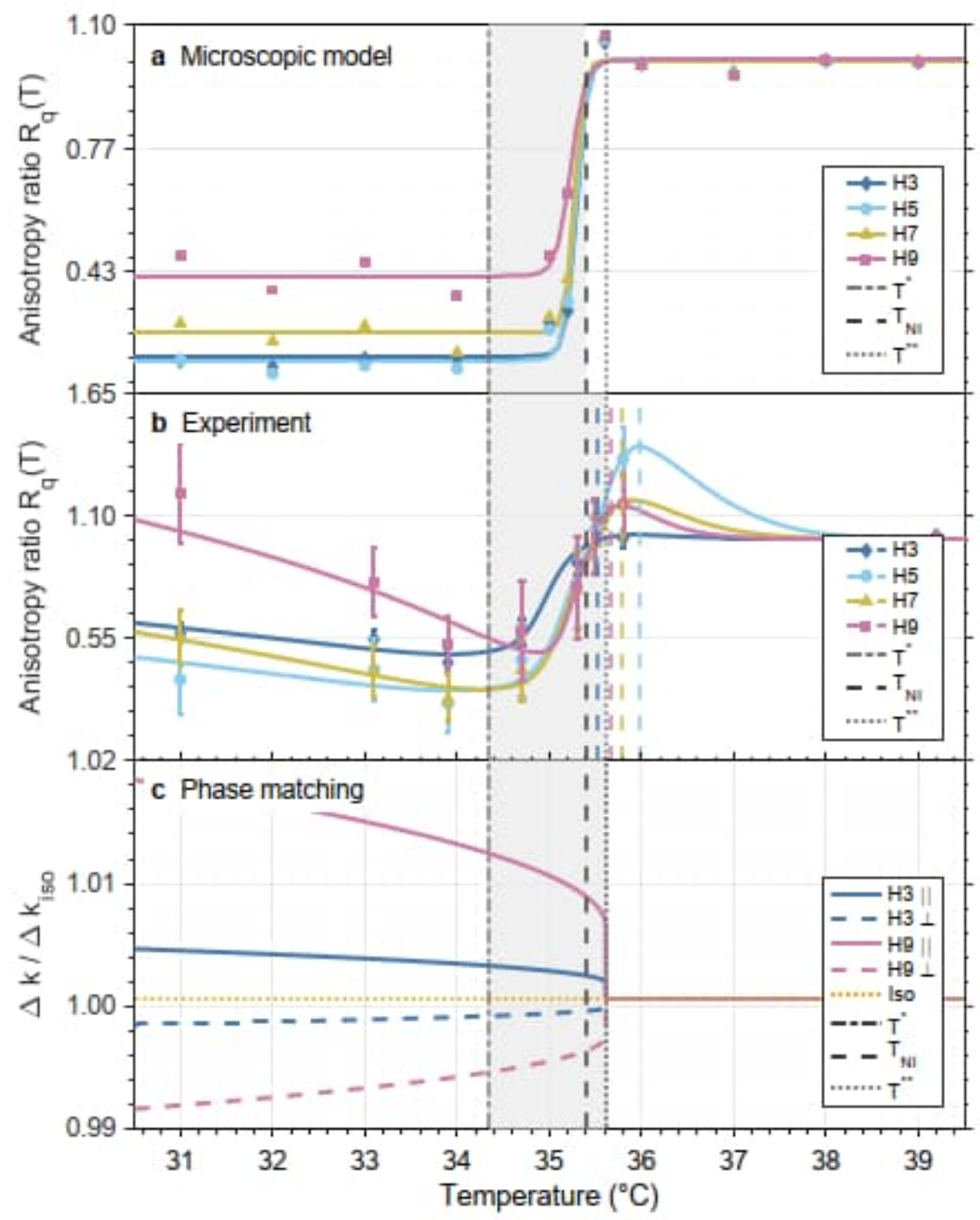

**Figure 4**: **Multiscale high-harmonic response across the** ***N–I*** **transition. a, TB model.** Temperature-dependent anisotropy ratio, $R_q(T)$, calculated for H3, H5, H7 and H9 using the TB model including intermolecular coupling. Symbols denote the calculated values, and solid curves show sigmoidal fits. The grey dash-dotted, black dashed and grey dotted vertical lines indicate the isotropic supercooling limit $T^*$, the nominal transition temperature $T_{\mathrm{NI}}$ and the nematic superheating limit $T^{**}$, respectively. The shaded region marks the interval $T^* < T < T^{**}$. **b, Experiment.** Experimental anisotropy ratios $R_q(T)$ extracted from the polarization-resolved angular distributions. Symbols and error bars denote the measured values, and solid curves show the macroscopic Landau–de Gennes $Q$-tensor model fitted to the experimental data and constrained by the TB response in **a**. Coloured dashed vertical lines indicate the fitted optical-response midpoint $T_q$ for each harmonic order. The thermodynamic limits and shaded region are defined as in **a**. **c, Phase matching.** Calculated phase mismatch $\Delta k_{q,\alpha}(T)$, normalized to its isotropic value $\Delta k_{q,\mathrm{iso}}$, for H3 (blue) and H9 (magenta). Solid and dashed curves correspond to driving-field polarizations parallel and perpendicular to the nematic director, respectively. The coloured dotted horizontal line denotes the isotropic reference, $\Delta k_{q,\alpha}/\Delta k_{q,\mathrm{iso}} = 1$. The vertical lines and shaded region are defined as in **a**. All panels share the temperature axis shown below **c**.

Within the $Q$-tensor formalism developed in Supplementary Section 5.4, the HHG intensity in the nematic phase can be decomposed into three contributions for each harmonic order $q$ and for driving-field polarization parallel or perpendicular to the director $\hat{\mathbf{n}}$. The first contribution is coherent emission associated with the average molecular alignment. Its field amplitude scales with the order parameter $S(T)$, while the resulting intensity is modulated by a phase-matching factor determined by the jet thickness and the harmonic coherence length. The harmonic anisotropy is not treated as an independent fitting parameter but is constrained by the TB dipoles, thereby linking the macroscopic response directly to the molecular electronic structure.

The second contribution originates from thermal fluctuations of the nematic order parameter. In the normal-incidence geometry used here, the harmonic signal is sensitive only to the transverse nematic correlation length, denoted $\xi_{\mathrm{nem},\perp}$, which characterizes the orientational fluctuations sampled along the propagation direction. The third contribution arises from long-wavelength director fluctuations governed by the Frank elastic constants. This contribution provides a weaker, geometry-dependent background.

Above the transition, the order parameter vanishes, and the HHG intensity is governed by two contributions. The first is residual coherent emission associated with the boundaries of transient paranematic swarms and remains influenced by phase matching. The second is incoherent emission from fluctuations within these swarms, governed by the isotropic correlation length $\xi_{\mathrm{iso}}$. As $T_{\mathrm{NI}}$ is approached from above, $\xi_{\mathrm{iso}}$ increases, causing the incoherent contribution to dominate.

For each harmonic order, the model in Eq. (3), with the sigmoidal weight defined in Eq. (4), is fitted to the measured temperature dependence of $R_q(T)$. The adjustable parameters are the amplitudes of the mean-field and fluctuation contributions, the offset between the midpoint of the optical response and the thermodynamic transition temperature, and the crossover width $\Delta T_q$. The thermodynamic transition temperature $T_{\mathrm{NI}}$, the spinodal limits $T^*$ and $T^{**}$, the Landau–de Gennes coefficients, the Frank elastic constants, and the refractive indices at the fundamental and harmonic wavelengths are fixed using literature values or independent measurements. The temperature- and polarization-dependent phase mismatch is calculated using the Vuks relations (*59, 60*).

Further details of the model, parameter constraints, and fitting procedure are provided in Supplementary Section 5.4.

As shown in Figure 4b (solid lines), the $Q$-tensor model reproduces the measured temperature dependence of $R_q(T)$ for all harmonic orders across the full investigated range, from the nematic phase at 31.0 °C to the isotropic phase at 39.0 °C.

The decrease in $R_q(T)$ observed for all four harmonics upon heating through the nematic phase arises primarily from the temperature dependence of the scalar order parameter $S(T)$, whereas nematic fluctuations make only a minor contribution. The decrease as the transition is approached from the nematic side is therefore driven mainly by the reduction of the mean-field emission anisotropy between the parallel and perpendicular polarization channels. This macroscopic behaviour reflects the microscopic anisotropy of the nonlinear response associated with each harmonic order, as detailed in the Supplementary Information.

As follows from the definition of $R_q(T)$ in Eq. (1), the decrease in $S(T)$ as the transition is approached from the nematic side reduces the parallel contribution in the numerator while increasing the perpendicular contribution in the denominator, thereby lowering $R_q(T)$. This decrease occurs in the same nematic-side temperature region in which the parallel channel remains more strongly phase mismatched than the perpendicular channel, as shown in Figure 4c. The parallel emission is therefore suppressed more strongly, reinforcing the decrease observed in Figure 4b.

Closer to the transition, within $T^* < T < T_{\mathrm{NI}}$, the nematic correlation length $\xi_{\mathrm{nem}}$ increases but remains finite because the first-order transition occurs before the nematic spinodal limit $T^{**}$ is reached, as illustrated in Figure 1d (*35*). The contribution of nematic fluctuations to the harmonic signal is controlled by the ratio of their correlation length to the harmonic coherence length, $L_{\mathrm{coh},q,\alpha} = \pi/|\Delta k_{q,\alpha}|$, where $\alpha = \|, \perp$ denotes the driving-field polarization relative to the nematic director. As $|\Delta k_{q,\alpha}|\xi_{\mathrm{nem}}$ approaches unity, the fluctuation contribution crosses over from a regime that scales approximately as $\xi^2_{\mathrm{nem}}$ to one limited by $L^2_{\mathrm{coh},q,\alpha} \propto |\Delta k_{q,\alpha}|^{-2}$. Because the parallel and perpendicular channels experience different phase mismatches, as shown in Figure 4c, this crossover modifies their propagation efficiencies unequally and contributes to the further suppression of $R_q(T)$ near the transition observed in Figure 4b.

As the temperature increases through $T_{\mathrm{NI}}$, the polarization ratio $R_q(T)$ exhibits a characteristic overshoot above unity. Within the $Q$-tensor model, the overshoot is attributed to emission associated with paranematic swarms. Physically, upon entering the isotropic phase, the sudden disappearance of macroscopic birefringence abruptly lifts the severe phase-matching penalty that had previously suppressed the parallel emission. Because these paranematic swarms retain a statistical orientational bias along the former nematic director, the parallel channel remains more efficient than the perpendicular channel despite the loss of macroscopic order. This orientational memory persists until the isotropic correlation length $\xi_{\mathrm{iso}}$ decreases at higher temperatures (*35, 61*).

While the heightened sensitivity of higher-order nonlinear processes to local fluctuations generally enhances this near-transition response, the exact magnitude of the overshoot is determined by the interplay between the microscopic nonlinear susceptibility and the local phase-matching constraints. Consequently, both our model calculations and experimental data show that H5 exhibits the most prominent overshoot among the investigated harmonic orders.

The particularly pronounced H5 overshoot arises because the birefringence-induced suppression of the parallel channel, which depends on the jet thickness, is strongest for the fifth harmonic. This produces the deepest minimum in $R_5(T)$ on the nematic side of the transition. As the sample is heated through $T_{\mathrm{NI}}$ and the macroscopic birefringence vanishes, both polarization channels recover their propagation efficiency. Because the H5 parallel channel recovers from the strongest suppression, it undergoes the largest relative increase in efficiency and consequently exhibits the steepest recovery across the transition. Combined with the additional emission from paranematic swarms, which preferentially enhances the parallel channel, this rapid phase-matching recovery drives $R_5(T)$ significantly above unity.

Deep in the isotropic phase, $\xi_{\mathrm{iso}}$ becomes comparable to the bare intermolecular distance $\xi_0$, and the paranematic swarms disappear. In this limit, the medium becomes fully disordered and $R_q$ approaches unity.

**Table 1**: **Fitted harmonic-response midpoints.** Values of $T_q$ extracted from least-squares fits of the anisotropy ratio $R_q(T)$ for each harmonic order. The uncertainties are obtained from the covariance matrix of the fits. The overlapping intervals indicate that the fitted midpoints are consistent with a common thermodynamic transition temperature.

| | H3 | H5 | H7 | H9 |
|---|---|---|---|---|
| $T_q$ (°C) | 35.5 ± 0.2 | 36.0 ± 0.5 | 35.8 ± 0.3 | 35.7 ± 0.3 |

Crucially, H9 exhibits a significantly steeper temperature derivative of the phase mismatch ($d\Delta k/dT$) and a far more pronounced variation in $L_{\mathrm{coh}}$ compared to H3. This harmonic-scaling effect acts as a macroscopic amplifier of the subtle thermal variations in the order parameter, which directly modulate the medium's dispersion, rendering the highest harmonic orders uniquely sensitive to the immediate onset of the phase transition. Consequently, our framework captures a systematic progression in effective transition temperatures rather than separate thermodynamic thresholds (see Table 1). This arises because each harmonic order acts as a distinct nonlinear filter, probing the spatial evolution of the phase transition with higher precision as the order increases.

# Discussion

We have investigated HHG in a flow-aligned LC by combining experiments with multi-scale theoretical and computational modelling. Using a temperature-controlled flat jet, we measured polarization-resolved HHG across the $N$–$I$ phase transition of 5CB. In the nematic phase, the emission exhibits a four-lobed anisotropy dominated by emission for driving-field polarization perpendicular to the director. DFT calculations and semi-empirical TB models attribute this response to intermolecular orbital hybridization within $\pi$-stacked antiparallel dimers. To bridge the microscopic and mesoscopic regimes, we developed a Landau–de Gennes $Q$-tensor model that incorporates phase-matched optical wave propagation. This framework establishes that the temperature- and harmonic-order-dependent polarization anisotropy ratio $R_q(T)$, defined as the har-monic yield for driving-field polarization parallel to the director relative to that for perpendicular polarization, is governed by the interplay between the scalar order parameter, birefringence-induced phase mismatch, and thermal orientational fluctuations. In particular, the sharp overshoot of $R_q(T)$ above unity immediately above $T_{\mathrm{NI}}$ is attributed to emission associated with short-range, transient paranematic swarms, an interpretation supported by the agreement between the experimental data and theoretical predictions across all observed harmonic orders.

More broadly, our results show that high-harmonic spectroscopy in anisotropic liquids can probe collective electronic responses and mesoscopic structural transitions beyond the single-molecule limit. By harnessing harmonic-dependent dispersion and phase mismatch, tracking HHG yields across phase boundaries offers an all-optical spectroscopic approach for mapping orientational order and pretransitional orientational fluctuations in soft matter. Leveraging the intrinsic attosecond time resolution of HHG, these findings pave the way for novel ultrafast spectroscopies capable of capturing real-time phase coexistence, domain-wall dynamics, and field-induced director reorientation in complex fluids. Furthermore, the enhanced sensitivity found in higher harmonic orders demonstrates that high-order optical nonlinearities function as spatiotemporal filters, amplifying subtle changes in local dispersion and order parameters. Although this study focused on a prototypical nematic LC, the underlying framework should readily extend to other soft matter systems, including smectic mesophases, lyotropic LCs and photoresponsive LC elastomers. This work establishes high-harmonic spectroscopy as a powerful paradigm for investigating multiscale structure and ultrafast dynamics in ordered liquid phases.

## Materials and Methods

### Experimental methodology

Mid-infrared pulses centred at 3.9 $\mu$m with a full-width-at-half-maximum duration of 130 fs were focused at normal incidence onto a continuously recirculating 5CB flat jet. The liquid sheet, with a thickness of 23.1 ± 0.2 $\mu$m, was operated at a flow rate of $30\,\mathrm{mL\,min^{-1}}$, providing a continuously refreshed interaction region. The driving-field intensity and polarization were controlled using wire-grid polarizers and motorized wave plates.

Temperature-dependent measurements were performed during heating across the $N$–$I$ transition. The temperature was measured after equilibration immediately before each acquisition and remained stable within $\pm 0.1\,^{\circ}\mathrm{C}$. The local mesophase and director configuration were independently verified by *in situ* polarization imaging. H3 was recorded with a near-infrared spectrometer, and H5–H9 were recorded with an ultraviolet–visible spectrometer. The wavelength- and polarization-dependent detection response was calibrated using a broadband source and an isotropic-phase reference. Full experimental, calibration and data-processing details are provided in the Supplementary Information.

**Theoretical methodology**

We modelled the high-harmonic response of 5CB using a two-dimensional semi-empirical TB framework adapted from models of molecular crystals (*23, 45*). Each molecule was represented by a two-site rod, and temperature-dependent structural ensembles were generated using the density $\rho(T)$ and orientational order parameter $S(T)$ of 5CB. Molecular orientations were sampled from a von Mises distribution, while anisotropic intermolecular packing and centre-of-mass disorder were included to reproduce the short-range structure of the LC. Intramolecular and nearest-neighbour intermolecular couplings were parameterized using first-principles calculations of the 5CB monomer and antiparallel dimer, with the distance dependence of the intermolecular hopping described by Harrison scaling. The electronic dynamics were propagated using the Peierls substitution for a pulse with a full-width-at-half-maximum duration of $130\,\mathrm{fs}$ centred at $3.9\,\mu\mathrm{m}$, and the harmonic spectra were obtained from the time-dependent electronic current. Temperature-dependent responses parallel and perpendicular to the nematic director were calculated from structural ensembles spanning the nematic and isotropic regimes. These microscopic responses were subsequently used to constrain the macroscopic $Q$-tensor model. Full computational details are provided in Supplementary Section 5.1.

## Acknowledgements

The authors thank Yannick Steinhauser, Michael Weiß Mare and Thomas Wagner, together with the staff of the mechanical workshop at the Max Planck Institute for Nuclear Physics in Heidelberg, for their technical expertise and contributions to the original design and continued development of the LC flat jet, including structural and temperature-control improvements implemented for this work.

## Author contributions

L.C. conceived the project, secured funding, provided resources, administered the project and supervised the research. M.L.M.-S. and A.A. developed the experimental methodology, designed and built the experimental setup, performed the experiments and developed the data-processing methodology. M.L.M.-S. prepared and handled the samples. M.L.M.-S. and A.A. analysed the experimental data. A.A. developed the macroscopic HHG Q-tensor formalism and the data-acquisition software. S.V.B.J. developed the theoretical model and performed the simulations under the supervision of A.R. and N.T.-D. M.L.M.-S. designed and prepared the figures. M.L.M.-S., A.A., S.V.B.J., A.R., C.V., T.P., D.F., N.T.-D., and L.C. validated and interpreted the results. M.L.M.-S., A.A. and L.C. wrote the initial manuscript. M.L.M.-S., A.A., S.V.B.J., A.R., C.V., T.P., D.F., N.T.-D., and L.C. reviewed and edited the manuscript. M.L.M.-S. and A.A. contributed equally.

## Funding

This work was supported by the European Research Council Synergy Grant UnMySt (ERC-2024-SyG-101167294). Additional support was provided by the Cluster of Excellence Advanced Imaging of Matter (AIM), the Gobierno Vasco Grupos Consolidados y de Alto Rendimiento programme (IT1453-22), and the Max Planck-New York City Center for Non-Equilibrium Quantum Phenomena. The Flatiron Institute is a division of the Simons Foundation. S.V.B.J. acknowledges support from the Alexander von Humboldt Foundation.

## Competing interests

The authors declare no competing interests.

## Data availability

All data needed to evaluate the conclusions in the paper are present in the paper and/or the Supplementary Materials.

## Code availability

The electronic-structure calculations were performed using the open-source Octopus package. The custom code used for data processing, angular fitting, tight-binding and $Q$-tensor modelling, model fitting, and figure generation is available from the corresponding authors upon reasonable request.